\documentclass[longbibliography,aps,prb,twocolumn,superscriptaddress,showpacs,amscd,amsmath,amssymb,verbatim]{revtex4-2}

\usepackage{graphicx}
\usepackage[OT1,T1]{fontenc}
\usepackage{revsymb}
\usepackage{bm}
\usepackage[dvipsnames]{xcolor}
\usepackage[separate-uncertainty=true,per-mode=power,inter-unit-product =\cdot]{siunitx}
\usepackage[colorlinks=true,linkcolor=blue,citecolor=blue,urlcolor=blue]{hyperref}
\usepackage{soul}
\usepackage{lineno}
\usepackage{placeins}
\usepackage{stmaryrd}

\newcommand{\TTO}{Tb$_2$Ti$_2$O$_7$}
\newcommand{\HTO}{Ho$_2$Ti$_2$O$_7$}
\newcommand{\DTO}{Dy$_2$Ti$_2$O$_7$}
\newcommand{\TTOx}{Tb$_{2+\mathrm{x}}$Ti$_{2-\mathrm{x}}$O$_{7+\mathrm{y}}$}
\newcommand\mrm[1]{\mathrm{#1}}

\DeclareSIUnit \mub {\ensuremath{\mu_{B}}}

\makeatletter
\def\maketitle{
\@author@finish
\title@column\titleblock@produce
\suppressfloats[t]}
\makeatother

\begin{document}

\title{Exploring possible magnetic monopoles-induced magneto-electricity in spin ices}

\author{Y. Alexanian}
\email[]{Yann.Alexanian@unige.ch}
\altaffiliation[]{current address: Department of Quantum Matter Physics, University of Geneva, 24 Quai Ernest-Ansermet, CH-1211, Geneva, Switzerland}
\affiliation{Institut Néel, CNRS and Université Grenoble Alpes, BP166, F-38042 Grenoble Cedex 9, France}

\author{J. Saugnier}
\affiliation{Institut Néel, CNRS and Université Grenoble Alpes, BP166, F-38042 Grenoble Cedex 9, France}

\author{C. Decorse}
\affiliation{ICMMO, Université Paris-Saclay, CNRS, 91400 Orsay, France}

\author{J. Robert}
\affiliation{Institut Néel, CNRS and Université Grenoble Alpes, BP166, F-38042 Grenoble Cedex 9, France}

\author {R. Ballou}
\affiliation{Institut Néel, CNRS and Université Grenoble Alpes, BP166, F-38042 Grenoble Cedex 9, France}

\author{E. Lhotel}
\affiliation{Institut Néel, CNRS and Université Grenoble Alpes, BP166, F-38042 Grenoble Cedex 9, France}

\author{J. Debray}
\affiliation{Institut Néel, CNRS and Université Grenoble Alpes, BP166, F-38042 Grenoble Cedex 9, France}

\author{F. Gay}
\affiliation{Institut Néel, CNRS and Université Grenoble Alpes, BP166, F-38042 Grenoble Cedex 9, France}

\author {V. Simonet}
\affiliation{Institut Néel, CNRS and Université Grenoble Alpes, BP166, F-38042 Grenoble Cedex 9, France}

\author {S. de Brion}
\email[]{sophie.debrion@neel.cnrs.fr}
\affiliation{Institut Néel, CNRS and Université Grenoble Alpes, BP166, F-38042 Grenoble Cedex 9, France}

\date{\today}

\begin{abstract} The possibilities of combining several degrees of freedom inside a unique material have recently been highlighted in their dynamics and proposed as information carriers in quantum devices where their cross-manipulation by external parameters such as electric and magnetic fields could enhance their functionalities. An emblematic example is that of electromagnons, spin-waves dressed with electric dipoles, that are fingerprints of multiferroics. Point-like objects have also been identified, which may take the form of excited quasiparticles. This is the case for magnetic monopoles, the exotic excitations of spin ices, that have been recently proposed to carry an electric dipole although experimental evidences remain elusive. Presently, we investigate the electrical signature of a classical spin ice and a related compound that supports quantum fluctuations. Our in-depth study clearly attributes magnetoelectricity to the correlated spin ice phase distinguishing it from extrinsic and single-ion effects. Our calculations show that the proposed model conferring magnetoelectricity to monopoles is not sufficient, calling for higher order contributions.
\end{abstract}

\maketitle

\section*{Introduction \label{sec:intro}}
Behind their strong interest for the stabilization of original magnetic states, magnetic frustrated systems provide a wealth of emergent excitations, which are prone to couple with multiple degrees of freedom. Among these, emergent magnetic monopoles in spin ices have been proposed by D. I. Khomskii to carry an electric dipole \cite{Khomskii2012} opening the route towards new magneto-electric effects and multiferroicity in spin ices and related phases.

\begin{figure*}
\centering
\includegraphics[width=\textwidth]{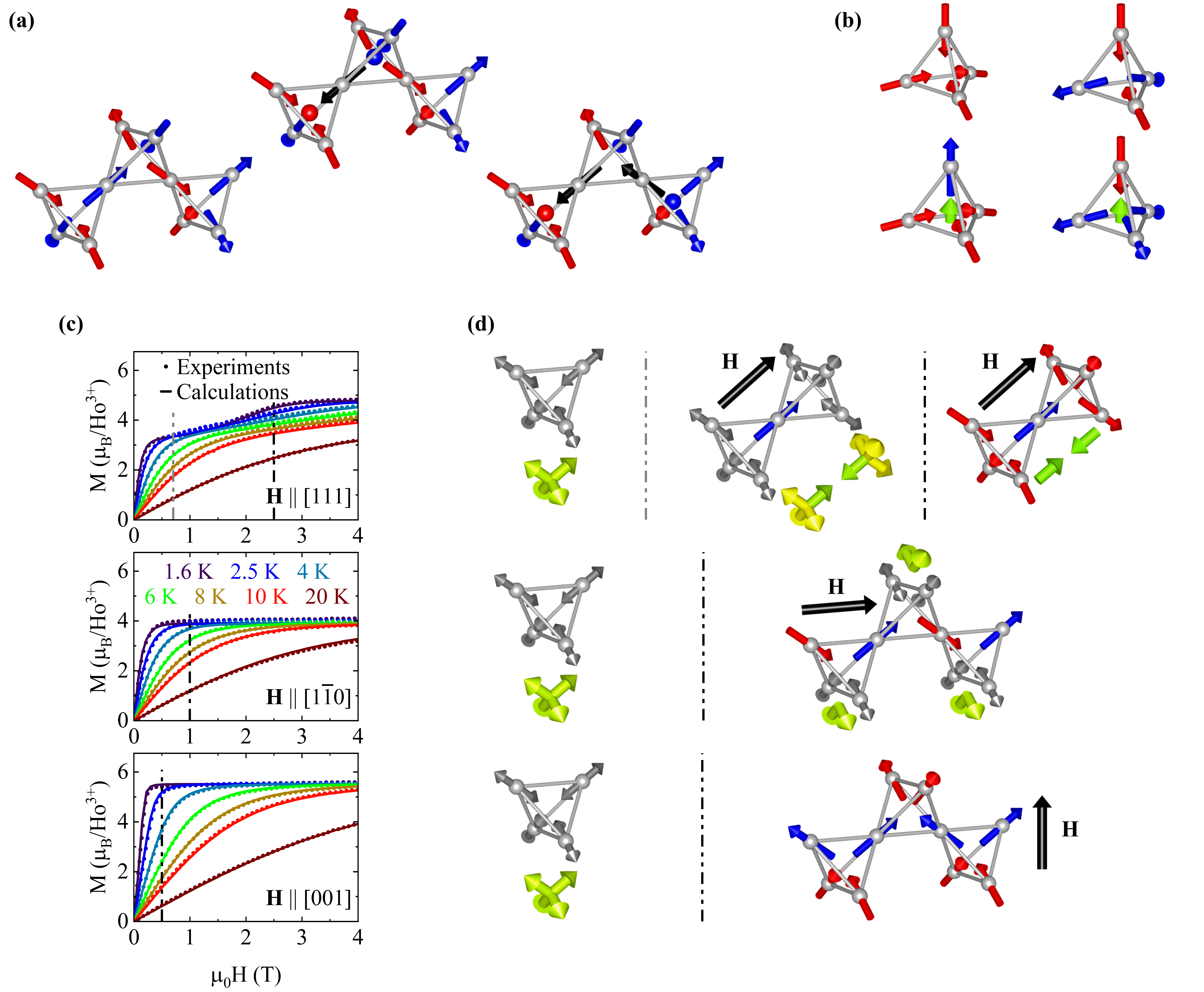}
\caption{{\bf Magnetic and electric configurations in spin ice and its field induced phases}. Tetrahedra in the pyrochlore lattice are depicted in grey. Red, blue and black arrows represent the spins oriented along a local $[111]$ direction. The grey double head arrows feature a degeneracy in the orientation of the spins due to thermal fluctuations. Yellow to green arrows indicate the possible directions of electric dipole moments associated to one tetrahedron in Khomskii’s model \cite{Khomskii2012}, from the least to the most likely at low temperature (typically 2.5 K).
(a) Spin ice phase with the two-in two-out ice rule on each tetrahedron (left) with a pair of monopoles (middle) and their deconfinement (right). (b) Prediction of electric dipole depending on the spin configuration in one tetrahedron. (c) Experimental (dots) and calculated (lines) magnetization curves of \HTO\ at several temperatures for different orientations of the magnetic field. The vertical dashed lines indicate the changes in the calculated permittivity (displayed in Fig. \ref{fig:LowT}) associated to the different phases at 2.5 K. The corresponding  spin arrangements at temperatures allowing the presence of monopoles are schematized in (d). 
}
\label{fig:Diagram}
\end{figure*}

The spin ice state is realized when magnetic moments on a \ lattice (featuring corner sharing tetrahedra) have an Ising anisotropy along the local $\langle 111 \rangle$ directions  and interact through effective ferromagnetic interactions $J$ \cite{Harris1997} (see Fig. \ref{fig:Diagram}). It is a macroscopically degenerate state in which spins obey a local constraint:  on each tetrahedron, two spins point inward and the other two outward (the so-called ice rule, in reference to water ice). The  magnetic excitations that emerge from this spin ice state can be described as magnetic charges (monopoles) located at the centers of the tetrahedra and which correspond to a violation of the local ice rule \cite{Ryzhkin2005,Castelnovo2008}. In the pyrochlore lattice, monopoles are created in pairs of opposite charges that can move apart. 

\begin{figure}[h!]
\includegraphics[width=1\columnwidth]{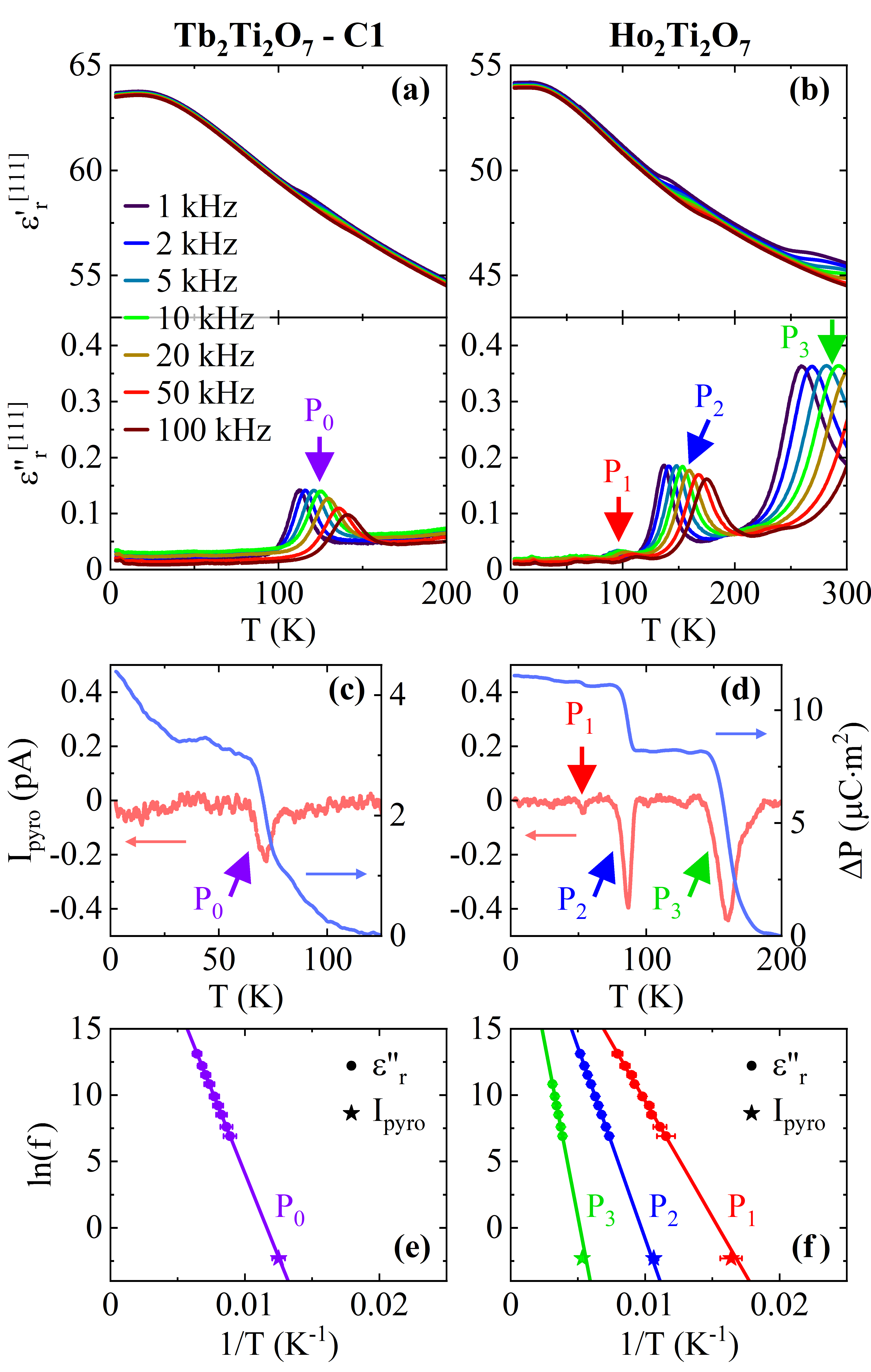}
\caption{{\bf Temperature dependence of the electric properties of \TTO\ (left) and \HTO\ (right) in zero magnetic field.} (a-b) Real part $\varepsilon'_{\mrm{r}}$ (top) and imaginary part $\varepsilon''_{\mrm{r}}$ (bottom) of the permittivity measured at different frequencies from $\qty{1}{\kilo\hertz}$ to $\qty{100}{\kilo\hertz}$. Dissipation peaks $P_{0}$ to $P_{3}$ observed in $\varepsilon''_{\mrm{r}}$ are associated with a frequency dependent anomaly in $\varepsilon'_{\mrm{r}}$. (c-d) Pyroelectric current measured at a speed of $v = \qty{4}{\kelvin/\minute}$ and corresponding change of the electric polarization using a reference at $\SI{150}{\kelvin}$ (for \TTO) and $\SI{200}{\kelvin}$ (for \HTO). (e-f) Arrhenius plots associated with these $\varepsilon''_{\mrm{r}}$ dissipation peaks: the measurement frequency is reported in a semi log plot as a function of the peak position in inverse temperature (dots). The straight lines are linear fits from which are extracted the activation energy and zero-point relaxation time. For the DC pyroelectric measurements (stars), we use a characteristic time of $\SI{10}{\second}$ corresponding to a frequency of $\SI{0.1}{\hertz}$ and the temperature was chosen at the initial rise of the pyroelectric current.}
\label{fig:HighT}
\end{figure}

Magnetic monopoles have been successfully probed in the prototype \DTO\ and \HTO\ spin ice compounds using various techniques from neutron scattering to magnetic noise measurements \cite{Dusad19, Bramwell2020, Jaubert2021}. 
It is worth noting that the spin ice ground state in zero magnetic field being a vacuum of monopoles, finite temperature and / or magnetic field are required to generate a finite density of monopoles. Typically in \HTO\ where $J\sim 1.8$ K, a significant density of magnetic monopoles is expected for temperatures larger than 2 K \cite{denHertog2000, Jaubert2011}.
In the presence of a magnetic field $\bm{H}$, the degenerate spin ice phase is gradually suppressed. It gives rise to a monopole crystal when the field is applied along the $[111]$ direction, a partially disordered phase for ${\bm H}\parallel [110]$, and an ordered spin ice phase for ${\bm H}\parallel [001]$ (see Fig. \ref{fig:Diagram}) \cite{Harris1998}. Applying a magnetic field along these peculiar directions thus appears as a control parameter of choice to probe the possible magneto-electric effects associated to magnetic monopoles. 

Magneto-electric effects were  observed in \HTO\ \cite{Katsufuji2004} and \DTO\  \cite{Saito2005,Grams2014} at low temperatures, for field applied along the $[001]$ and $[111]$ directions. In the latter field direction, the dielectric dynamics were proposed to have a critical behavior associated to monopole condensation \cite{Grams2014}. Sample dependent weak signatures of ferroelectricity were also reported \cite{Dong2009,Liu2013,Lin2015} and tentatively related to magnetic monopoles. The \TTO\ pyrochlore compound, which does not present a classical spin ice behavior, rather a still debated quantum spin liquid one \cite{Rau2019}, was also proposed theoretically to host magneto-electric monopoles in applied field \cite{Jaubert2015}, whose observation was claimed in subsequent experimental studies \cite{SanthoshKumar2021,Jin2020}. 

To obtain a unified experimental picture of these complex and sometimes contradictory results, we have performed an extensive dielectric and magneto-electric study by electric polarization and permittivity measurements on several single crystals of the classical spin ice \HTO\ and the quantum spin liquid candidate \TTO. The full $(H,T)$ phase diagram was investigated in the three main cubic directions ($[111]$, $[110]$ and $[001]$) in the 2.5 to 300 K temperature range. 
We show that, indeed, electric effects are present, with three different origins. The first one is extrinsic, sample-dependent, and associated to point defects. The second one, of magneto-dielectric character, is related to the individual rare earth magneto-dielectric response in cubic symmetry. The third one, only present in the Ho compound, is clearly associated with spin ice physics and correlated to the different phases of the $(H,T)$ spin ice  phase diagram. Our Monte-Carlo simulations based on the electric dressing of monopoles within Khomskii's model nevertheless show that these alone cannot explain the magneto-dielectric behavior of \HTO\ and that more complex ingredients must be considered. These results provide a consistent picture of magneto-electric effects in the investigated pyrochlore oxides and enlighten different contributions not necessarily related to magnetic monopoles.

\section*{Results}
\vspace{1\baselineskip}
{\bf Extrinsic electric response in the high temperature regime.}
The complex permittivity  for the Ho and Tb pyrochlores is shown in Fig \ref{fig:HighT}a,b up to $\SI{200}{\kelvin}$ and $\SI{300}{\kelvin}$ respectively. The two compounds behave similarly: $\varepsilon'_{\mrm{r}}$ hardens down to $\qty{30}{\kelvin}$ by $\qtyrange{15}{20}{\percent}$ and then saturates. 
In addition, frequency dependent dissipation peaks are visible in $\varepsilon"_{\mrm{r}}$ associated with small anomalies in $\varepsilon'_{\mrm{r}}$. The two compounds also present at the same temperatures small pyroelectric current peaks (and therefore changes of the electric polarization of the order of $\SIrange{1}{10}{\micro\coulomb \meter^{-2}}$) as seen in Fig. \ref{fig:HighT}c. The frequency dependence of all the $\varepsilon''_{\mrm{r}}$ peak positions follows an Arrhenius law $f=f_0 \exp(-E_a/T)$ with an activation energy $E_{\mrm{a}}$ in the $\qtyrange{1500}{5000}{\kelvin}$ range and a zero-point relaxation time $\tau_{0}=1/(2\pi f_0)$ of the order of $10^{-13}$s, as visible in Fig. \ref{fig:HighT}e,f. These values are typical of defects relaxation in oxides and have been already observed in pyrochlores \cite{Kamba2002,Yadav2019}. Note that the measured small pyroelectric currents are a signature of the dynamics of these electric defects.

In \TTO, we tested the dependence of the electric response on defects concentration:  two additional crystals C2 and C3 were studied having a controlled and very small off-stoichiometry \TTOx\ with $\mathrm{x}$ equal to $+0.003$ and $-0.003$ respectively. We do see changes in the number of activated processes and the amplitude of the $\varepsilon''_{\mrm{r}}$ peaks. The larger amplitude in crystal C1 (reported in Fig. \ref{fig:HighT}) indicates a larger amount of defects in this sample (See Supplementary Note 1).This interpretation in terms of point defects is also supported by the absence of any clear dependence of these signals neither on the electric field orientation nor on a magnetic field up to $\SI{4}{\tesla}$ (see Supplementary Note 2 and 3).

\vspace{1\baselineskip}
{\bf Quadratic magneto-dielectric response in the intermediate temperature regime.}
\begin{figure*}
\includegraphics[width=1\textwidth]{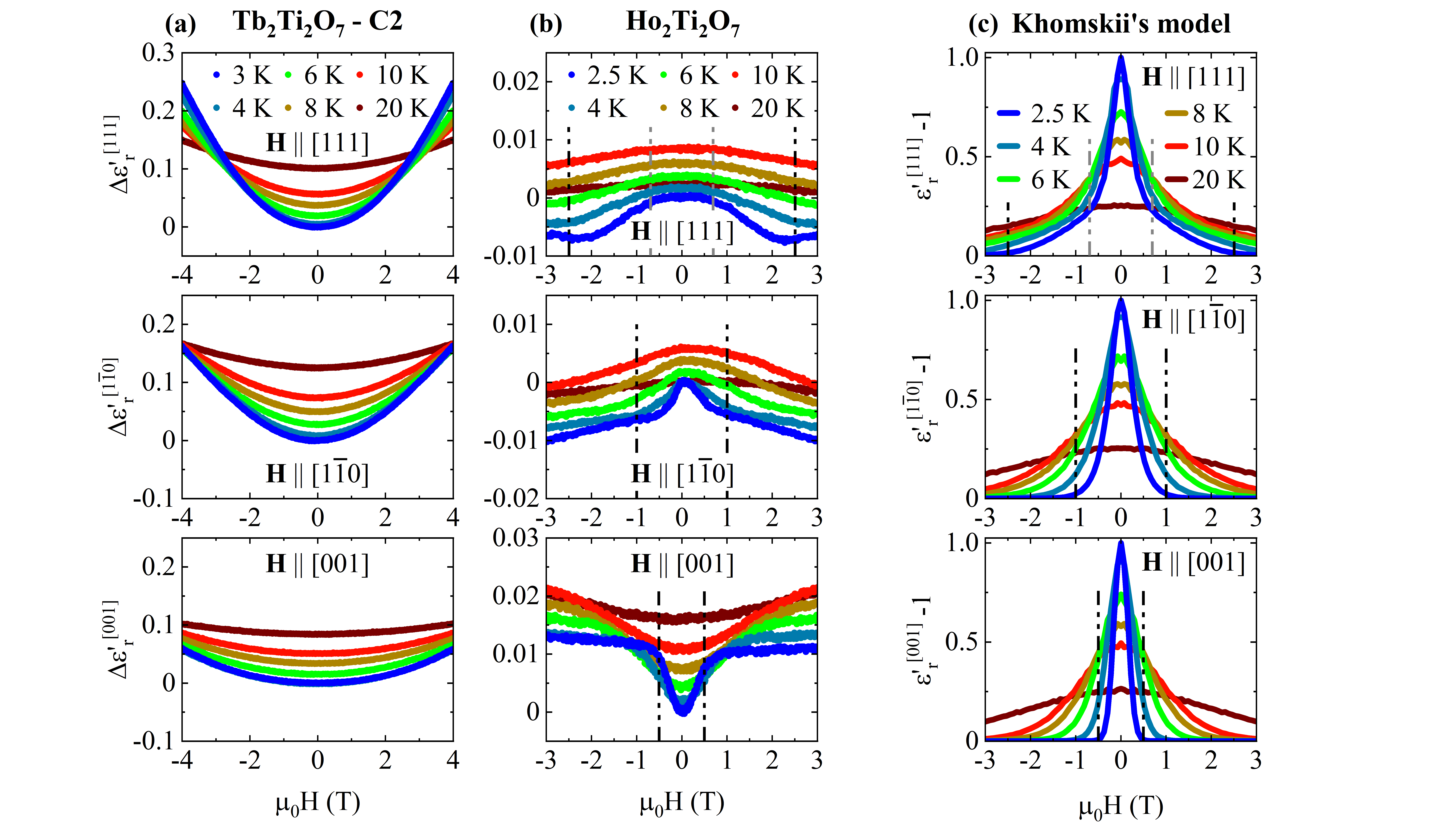}
\caption{{\bf Magneto-dielectric effects at low temperature in \TTO\ and \HTO\ compared to Khomskii’s model.} Magnetic field variation of the dielectric permittivity in in \TTO\ (a) and \HTO\ (b) below $\qty{30}{\kelvin}$ probed through the magnetic field variation of the permittivity with respect to its zero field value at $\qty{3}{\kelvin}$ and $\qty{2.5}{\kelvin}$ respectively. The electric field $\bm{e}$ is applied parallel to the static magnetic field $\bm{H}$ along the three main directions of the cubic pyrochlore structure. Note the different scales of $\Delta\varepsilon'_{\mrm{r}}$ for the two compounds. For \HTO, the applied magnetic field was corrected from the demagnetization field to allow a correct comparison with the spin ice phase diagram. (c) Monte-Carlo calculations of the dielectric susceptibility ($\varepsilon_{\mrm{r}}-1$) using Khomskii's model at the same temperatures and for the same field orientations as the measurements. It is normalized at 1 at $\qty{2.5}{\kelvin}$ and zero field. The vertical dashed lines in (c) point out to the changes of regime of the calculated dielectric susceptibility at $\SI{2.5}{\kelvin}$. These lines are reported on the \HTO\ measurements in panel (b) and on the magnetization measurements in Fig. \ref{fig:Diagram}c.}
\label{fig:LowT}
\end{figure*}
Figures \ref{fig:LowT}a,b give a view of the magneto-dielectric effects visible below $\qty{30}{\kelvin}$ for both \HTO\ and \TTO\ (crystal C2). All these measurements were done by applying an AC electric field $\bm{e}$ of frequency $f=\qty{10}{\kilo\hertz}$ fast enough to avoid interfacial charge effects, but still much slower than the characteristic frequency of spin relaxation in \HTO\ at 2.5 K ($f\sim 10^7$~Hz \cite{Ehlers2008})
so that the electric field can safely be considered as static compared to the monopole dynamics. 
Since no detectable effects are present in $\varepsilon''_{\mrm{r}}$, only $\varepsilon'_{\mrm{r}}$ is shown, as a function of the applied magnetic field in the same direction as the electric field. For both compounds, a quadratic dependence of $\varepsilon'_{\mrm{r}}$ with magnetic field is observed with a strength that increases when the temperature is lowered. Additional features superimposed on this quadratic behavior at the lowest temperatures for \HTO\ are discussed in the next section. 

Strikingly, the sign of this quadratic effect is opposite in the two compounds for $\bm{H}\parallel\bm{E}\parallel[111]$ and $\bm{H}\parallel\bm{E}\parallel[1\bar{1}0]$ and identical for $\bm{H}\parallel\bm{E}\parallel[001]$. It therefore depends on the compounds but also on the respective orientation of the electric and magnetic fields: The signs are globally inverted when ${\bm e} \perp {\bm H}$ (see Supplementary Information part D). The amplitude of the magneto-dielectric effect is also different: it is much larger in \TTO, where it scales with the $[111]$ component of the magnetic field. Finally, note that it is one order of magnitude larger than magneto-striction effects for both compounds \cite{Aleksandrov1985,Stöter2019}. These quadratic effects are reproducible as shown from measurements on the other \TTO\ samples (see Supplementary Note 4). They are attributed to the dielectric response of the isolated rare-earth ions as discussed later.

\vspace{1\baselineskip}
{\bf Low temperature behavior in \HTO\ and Monte-Carlo simulations of the Khomskii's model.}
Below $\qty{10}{\kelvin}$ and down to $\qty{2.5}{\kelvin}$, complex field dependencies are observed in \HTO\, that are not visible in \TTO. Anomalies in the permittivity appear for all directions of the magnetic field: around $\pm\SI{2}{\tesla}$ for $\bm{H}\parallel [111]$, around $\pm\SI{0.5}{\tesla}$ for $\bm{H}\parallel [1\bar{1}0]$ and around $\pm\SI{0.75}{\tesla}$ for $\bm{H}\parallel [001]$. Note the different changes in shape and curvature depending on the amplitude and direction of the magnetic field.
Magnetocapacitance measurements by Katsufuji {\it et al.} on a \HTO\ single crystal give similar results for $\bm{H}\parallel\bm{E}\parallel[001]$ and rather close ones for $\bm{H}\parallel\bm{E}\parallel[111]$ \cite{Katsufuji2004}.

Since these features appear concomitantly to spin ice correlations, it is tempting to relate them to magneto-electric monopoles dynamics and to the spin ice phase diagram under magnetic field. In the spin ice state, when a small magnetic field is applied along $[111]$, in a given tetrahedron, the spin whose easy axis is parallel to the field aligns with the field, the other three remaining in a degenerate state, the kagome ice. The ice rule is nevertheless preserved. At higher field, a crystal of monopoles is stabilized through a liquid-gas-like transition \cite{Sakakibara2003}. 

For $\bm{H}\parallel [1\bar{1}0]$, the system can be viewed as two orthogonal chains, resulting in the polarization of the spins along the magnetic field, while the others remain disordered but have to obey the ice rule \cite{Clancy2009}. For $\bm{H}\parallel [001]$, a topological transition, the Kasteleyn transition, occurs toward an ordered spin ice state, which maximizes the magnetization along the field \cite{Jaubert2008}. 
At finite temperature, these behaviors are smoothed due to thermal fluctuations, but the magnetization curves are reminiscent of these field induced transitions (see Fig. \ref{fig:Diagram}c,d). 

To test if the electric signatures of the magnetic phase diagram observed in \HTO\ are associated to the presence of monopoles, we have performed Monte Carlo simulations in the frame of Khomskii's model using a first neighbor spin Hamiltonian \cite{Bramwell2020} that reproduces correctly the compound magnetization in the temperature and magnetic field range studied (Fig. \ref{fig:Diagram}c). The expected magneto-dielectric contribution was computed assuming that each monopole carries an electric dipole along the appropriate $[111]$ direction (Fig. \ref{fig:Diagram}d). Note first that in the extended network of connected tetrahedra, the summed contribution of the local electric dipoles per tetrahedron may cancel resulting in a zero electric polarization. This is indeed what is calculated for all three directions of the magnetic field, in agreement with our measurements.

The permittivity calculated at 2.5 K is plotted in Figure \ref{fig:LowT}c. This quantity, related to the electric susceptibility, reflects the response of the electric dipoles to the modulated electric field  according to their correlations. These results have to be discussed with respect to the field induced behaviors recalled above. In zero magnetic field and at 2.5 K, spin ice correlations are present but thermal fluctuations are sufficient to produce monopole excitations and the associated local electric dipole moments, thus resulting in a non-zero dielectric permittivity. For $\bm{H}\parallel[111]$, the permittivity vanishes in two steps when increasing the absolute value of the magnetic field: first, additional constraints on the possible monopoles / electric dipoles in the kagome ice phase reduces the signal and second, the stabilized crystal of monopoles and antimonopoles results into anti-aligned electric dipoles with a zero contribution. For $\bm{H}\parallel[1\bar{1}0]$, the signal goes to zero, despite the fact that thermally excited monopoles can be present. This is attributed to the fact that the corresponding electric dipoles are oriented perpendicular to the exciting field $\bm{E}\parallel[1\bar{1}0]$. 
For $\bm{H}\parallel[001]$, the signal vanishes rapidly as expected since the saturated phase supports no magnetic monopole.


The magnetic fields corresponding to the changes of regime in the calculated Khomskii's electric susceptibility (indicated by vertical dashed lines in Fig. \ref{fig:LowT} and reported on the magnetization curves of Fig. \ref{fig:Diagram}c) indicate the field induced spin reorientations. They coincide quite well with the features observed in our permittivity measurements, establishing the electric sensitivity of \HTO\ to the magnetic phase diagram of its spin ice ground state dressed with monopoles. However the shape of the measured dielectric permittivity is not globally reproduced by the calculation, as discussed below.


\section*{Discussion}
We can then describe the electric effects in these pyrochlores with three different contributions. The first one, observable in Ho as well as in Tb compounds, is associated to defect relaxation processes and have no magnetic field dependence. These defects are probably related to Ti and O ions since they are present in both compounds. Similar activated processes were previously observed in \DTO\ and \HTO\ \cite{Dong2009,Yadav2019} with comparable activation energies. It should be emphasized that this defect contribution is certainly responsible for the small electric polarization measured previously in Dy, Ho and Tb compounds \cite{Dong2009,Liu2013,Lin2015,SanthoshKumar2021}: a few $\unit{\micro\coulomb.\meter^{-2}}$ in these systems \cite{Dong2009,Liu2013,Lin2015} to be compared to $\qty{6e5}{\micro\coulomb.\meter^{-2}}$ in the well known multiferroic BiFeO$_{3}$ \cite{Lebeugle2007}. We unambiguously show that these signals were therefore wrongly attributed to the monopole electrical contribution.  

\begin{table}[ht!]
\begin{ruledtabular}
\begin{tabular}{c|c|c}
$\delta (\SI{e-3}{\tesla^{-2}})$ & \hspace{1.3cm} Tb \hspace{1.3cm} & \hspace{1.3cm} Ho \hspace{1.3cm} \\
\hline\noalign{\vskip 0.5mm} 
$\delta_{xx}$ & $\num{1.1}$ & $\num{0.5}$ \\
$\delta_{xy}$ & $\sim \num{0}$ & $\sim \num{0}$ \\
$\delta_{44}$ & $\num{2.0}$ & $\num{-0.3}$ \\
\end{tabular}
\end{ruledtabular}
\caption{{\bf Quadratic contributions to the magneto-dielectric effect} at $T=\SI{20}{\kelvin}$ in both compounds. Note that $\delta_{xx}=\delta_{i=j=k=l}$ accounts for $\bm{E}\parallel\bm{H}$ processes, $\delta_{xy} = \delta_{i=j\ne k=l}$ for $\bm{E}\perp\bm{H}$ ones, and $\delta_{44}=\delta_{i\ne j,k\ne l}$ is related to mixed terms due to the fact that two components of $\bm{E}$ and $\bm{H}$ enters in Eq. \ref{eq:permittivity_tensor}. The other components of the $\delta$ tensor are zero by symmetry. The value given for $\delta_{xx}$ is obtained from the measurements with $\bm{E}\parallel\bm{H}\parallel[001]$ while $\delta_{44}$ is an average value deduced from measurements with four different $\bm{E}$ and $\bm{H}$ configurations.}
\label{tab:1}
\end{table}

The second contribution, quadratic with magnetic field, is visible mainly below $\SI{30}{\kelvin}$. It can be described phenomenologically through the quadratic magneto-dielectric tensor $\delta$ of these cubic pyrochlore systems. Indeed, the quadratic variation of $\varepsilon'_{\mrm{r}}$ can be expressed as
\begin{equation}
\Delta\varepsilon'_{\mrm{r}}(H) = (\mu_{0}H)^{2}\sum_{ijkl}\delta_{ijkl}e_{i}e_{j}h_{k}h_{l}
\label{eq:permittivity_tensor}
\end{equation}
where $e_{i,j}$ and $h_{k,l}$ are the normalized components of the applied low frequency AC electric field $\bm{E}$ and static magnetic field $\bm{H}$ respectively and $\delta_{ijkl}$ are the components of the quadratic magneto-dielectric tensor $\delta$. Due to the high symmetry of pyrochlore compounds, only few components are non zero. The magnetic space groups of the different magnetic phases in the spin ice phase diagram, as well as the corresponding usual reduced $6\times6$ matrix form of the $\delta$ tensor are given in the Supplementary Note 5. Assuming continuous and small field induced departures of the magnetic structure at $T\ge\qty{20}{\kelvin}$, we show that $\Delta\varepsilon'_{\mrm{r}}(H)$, in all our experimental configurations, can be expressed as a function of only 3 independent $\delta$ components, $\delta_{xx}$, $\delta_{xy}$ and $\delta_{44}$, of the paramagnetic space group $\mrm{Fd\bar{3}m1'}$ (see Table \ref{tab:1} and Supplementary Information part F). From our experimental results, we further deduce that for both compounds $\delta_{xy}(T)\approx 0$ and $\delta_{xx}(T) >0$ with a value two times larger in \TTO. Finally, ${\delta_{44}}(T)$ is positive for \TTO\ and negative for \HTO, with a magnitude seven times larger in \TTO\ (see Table \ref{tab:1}). 

This variability of behaviors according to the rare earth reveals a contribution of single-ion nature to the magneto-dielectric response functions. Possible mechanisms have been discussed in the framework of multiferroic rare-earth ferroborates \cite{Popov2013}. Our measurements  allow to quantify the magneto-dielectric response of these ions in the pyrochlore compounds, that is found much larger in Tb than in Ho compound. Interestingly, \TTO\ is also known to present large dynamical spin-lattice couplings already at high temperature \cite{Aleksandrov1985,Mamsurova1986,Nakanishi2011,Guitteny2013,Constable2017} that contribute to the unconventional spin liquid behavior reported at very low temperature. The observed giant magneto-dielectric effect suggests that these dynamical spin-lattice fluctuations are electrically active.

The last and most-original magneto-dielectric contribution, visible below $\SI{10}{\kelvin}$ only in \HTO, is strongly related to the spin ice magnetic phase diagram. The additional contributions to the quadratic dependency of $\varepsilon'_{\mrm{r}}$ are indeed observed at the phase boundaries between different spin arrangements for $\bm{H}\parallel[111]$ and $\bm{H}\parallel[1\bar{1}0]$. However, the Khomskii's model cannot account for some of the observed behavior as a function of magnetic field. For $\bm{H}\parallel[111]$, there is no peak centered at $H=0$ in the measurements contrary to the calculations. Instead, a slow decrease followed by a steeper one is observed up to $\pm2$~T. This is close to what is reported in the Dy$_2$Ti$_2$O$_7$ spin ice \cite{Grams2014,Saito2005}, with a signal around $\qty{1}{\tesla}$, broad at $\qty{2}{\kelvin}$ and narrowing down to $\qty{0.4}{\kelvin}$, below which it disappears. This signal was attributed to a speeding up of criticality and therefore a drastic change of monopole density around the liquid-gas transition  \cite{Grams2014}. Another discrepancy, even more obvious, occurs for $\bm{H}\parallel[001]$. There, we observe an increase in $\Delta\varepsilon'_{\mrm{r}}(H)$ while the density of monopoles is known to decrease and fall to zero since the magnetic field is stabilizing an ordered two-in two-out spin structure without monopoles.

Additional ingredients should therefore be considered. First of all, our calculations include the magnetic dipolar interactions through their first neighbor part only, giving an effective first neighbor interaction  $J=\SI{1.8}{\kelvin}$. The long-range part, which manifests into Coulomb interactions between monopoles \cite{Castelnovo2008}, is known to play a role in particular in its lowest temperature range \cite{Jaubert2011,Halln2022}. Dipolar electric interactions have not been included either, while they may stabilize ordered electric arrangement \cite{Jaubert2015}. However, all these dipolar effects should only alter slightly the monopole density and  and therefore its contribution to the dielectric permittivity and therefore cannot explain for instance the change of sign of the magneto-dielectric effect between our calculations and the experiments when $\bm{E}\parallel\bm{H}\parallel[001]$. We therefore seek for another contribution, especially large for $\bm{H}\parallel [001]$ when two-in two-out configurations are present, that would mask Khomskii’s contribution. Although neither exchange-striction nor spin currents are foreseen to generate electric dipoles in such symmetric local configurations, we performed additional calculations by imposing an arbitrary electric dipole proportional to the total magnetization of each tetrahedron. These were qualitatively able to reproduce the observed permittivity for $\bm{H}\parallel[001]$. However, no global agreement was achieved for other directions of the magnetic and electric fields. Additional calculations were attempted referring to several phenomenological proposals in the literature that associate the permittivity either to the nearest-neighbor spin correlations or to the nearest-neighbor tetrahedron magnetization correlations \cite{Saito2005}. Those models were also unsuccessful to account for our measurements performed for many configurations of $\bm{e}$ and $\bm{H}$, which therefore provide a very constrained set of data contrary to previous studies \cite{Saito2005,Grams2014,Katsufuji2004}. Note that the largest discrepancy between Khomskii's model and our measurements occurs for $\bm{H}\parallel[001]$, which could be related to the topological nature of the transition to the saturated state at low temperature along this direction \cite{Jaubert2008}.

The absence of a similar signal in \TTO\ could be due to the departure of \TTO\ from classical spin ice physics. The magnetic moments associated to the Tb$^{3+}$ ions have a much weaker Ising character allowing them to tilt in a magnetic field, thus altering the spin ice phase diagram. Actually, the enhanced quantum fluctuations in \TTO\ bring it closer to a quantum spin liquid. Khomskii's mechanism, conferring an electric dipole to any monopole configuration, was nevertheless predicted to occur in \TTO\ under magnetic field and to stabilize a bilayered crystal of monopoles \cite{Jaubert2015}. We did not observe any magneto-electric signature of this behavior, probably because these subtle effects are masked by the strong single-ion quadratic magneto-dielectric contribution in this compound. 


In conclusion, based on dielectric measurements, we have succeeded in disentangling the various contributions involved in the electrical response of pyrochlore compounds. In addition to extrinsic contributions at high temperature, we have isolated a quadratic single-ion response that seems particularly large in \TTO\ and could shed further light on the pathological behavior of this quantum spin liquid material. At the lowest temperature, in the canonical spin ice \HTO, we observed a clear signature in the permittivity of the spin-ice correlation regime, in all magnetic field directions. These subtle features were confronted with calculations assuming the dressing of magnetic monopoles by electric dipoles as proposed by Khomskii \cite{Khomskii2012}. The absence of full agreement between experiments and modelisation clearly shows that the direct observation of the monopole electric moments is masked by other stronger contributions. One of them, which has not been considered so far, is the rare earth single ion magneto-dielectric effect that we have shown to be present with a quadratic dependence on the magnetic field and an increased amplitude at low temperatures. This induced  electric moment should interact with the electrically dressed monopoles and may be responsible for the complex magneto-dielectric response of this spin ice compound.

\newpage

\section*{Methods} 

\HTO\ and \TTO\ single crystals were grown following the procedure described in Reference \cite{Gardner1998} with a slow growth rate of 2 mm/h. All crystals were then annealed in O$_2$ atmosphere to remove thermal stress and adjust the oxygen stoichiometry. \TTO\ samples C2 and C3 are those studied in Reference \cite{Alexanian2022}.
The magnetic and magneto-electric characterizations were performed on 3 different plaquettes to probe the electric response along the three main directions of the cubic pyrochlore structure: $[111]$, $[1\bar{1}0]$ and $[001]$. These plaquettes were cut from the same single crystal with typical dimensions $3\times3\times\qty{0.2}{\milli\meter^{3}}$.

Magnetization measurements were performed using a custom extraction magnetometer in the temperature range 1.5 K- 300 K using magnetic field up to 4 T. The plaquettes were aligned perpendicular to the magnetic field with an accuracy of a few degrees. Corrections for demagnetization fields were applied.

For magneto-dielectric measurements, silver paste was used as electrodes to constitute a parallel plate capacitor. The impedance was measured thanks to an impedance meter Agilent E4980A operating in the frequency range $f=\qtyrange{1}{500}{\kilo\hertz}$ with the applied AC electric field $\bm{e}$ perpendicular to the plaquette. Using a RC parallel circuit model, the real and imaginary parts of the sample (relative) permittivity (dielectric constant) $\varepsilon_{\mrm{r}}$ was deduced from the following formula: $C=\varepsilon_{0}\varepsilon'_{\mrm{r}}S/d$ and $1/R\omega=\varepsilon_{0}\varepsilon''_{\mrm{r}}S/d$ where $S$ is the plaquette surface, $d$ its thickness, $\varepsilon_{0}$ is the vacuum permittivity. Neglecting the magneto-striction and dilatation effects \cite{Aleksandrov1985,Stöter2019}, the temperature and magnetic field dependence of $C$ and $R$ are directly related to those of the permittivity $\Delta C/C \approx \Delta\varepsilon'_{\mrm{r}}/\varepsilon'_{\mrm{r}}$ and $\Delta R/R\approx-\Delta\varepsilon''_{\mrm{r}}/\varepsilon''_{\mrm{r}}$. A custom horizontal split-ring superconducting magnet allows to apply magnetic fields $\mu_{0}H$ up to $\qty{4}{\tesla}$ in different orientations with respect to the sample capacitor. Measurements with $\bm{E}\parallel\bm{H}$ and $\bm{E}\perp\bm{H}$ were performed in the temperature range $\qtyrange{2.5}{300}{\kelvin}$. To ensure a good comparison with the spin ice phase diagram of \HTO, corrections for demagnetization fields were applied for the \HTO\ plaquettes using a demagnetization coefficient of $0.9$ for $\bm{E}\parallel\bm{H}$ and $0.05$ for $\bm{E}\perp\bm{H}$. 

Pyroelectric measurements were performed on the same capacitor-like samples using a femto-electrometer Keithley 6517B. The samples were cooled down to $\qty{2}{\kelvin}$ with an electric bias of $\qty{200}{\volt}$. At $\qty{2}{\kelvin}$, the electric bias was removed and the temperature stabilized for $\qty{30}{\min}$ to ensure evacuation of the charges accumulated on the electrodes. The sample was then heated at a constant rate of $\qty{4}{\kelvin.\minute^{-1}}$ and the pyroelectric current measured as a function of time and then converted as a function of temperature. The polarization was obtained by integrating the pyroelectric current using a reference at high temperature.

The magnetization and permittivity curves have been calculated through Monte Carlo simulations, combining a single spin-flip Metropolis update with a loop algorithm \cite{Lefrancois2017}, on $16\times L^3$ lattice sites with $L = 8$ and periodic boundary conditions. The Hamiltonian includes an effective first neighbor ferromagnetic interaction $J=1.8$~K between Ising spins, and a Zeeman contribution accounting for the applied magnetic field. The dielectric susceptibility was computed over $10^5$ steps by evaluating the fluctuations of the electrical polarization following Khomskii’s model: an electric dipole was associated to each tetrahedron with a 3-in 1-out and 3-out 1-in configuration, the dipole being oriented from the center of the tetrahedron to the spin with a different orientation from the other three \cite{Khomskii2012}. The associated permittivity was then derived from the susceptibility. The calculated magnetization curves were normalized to the measurements using a magnetic moment of $\SI{9.55}{\mub}$ for the Ho$^{3+}$ ion (see Fig.~\ref{fig:Diagram}c).

\section*{Data availability}
The datasets used or analyzed during the current study are available from the corresponding authors on reasonable request.

\section*{Code availability}
The code used for this study is not publicly available but may be made available to qualified researchers on reasonable request from the corresponding author.

\section*{Authors contributions}
The project was initiated by S.dB. The single crystals were grown and characterized structurally by C.D.; they were shaped and oriented by Y.A. and J.D. Their magnetic properties were measured and analysed by
Y.A. and S.deB and their magneto-dielectric properties by Y.A., J. S., F.G. and S.dB. Monte Carlo simulations were performed by Y.A. and J.R. The results were discussed and then shaped into the manuscript by
Y.A., C.D., J.R., R. B., E. L., V.S. and S.dB.

\section*{Competing Interests}
The Authors declare no Competing Financial or Non-Financial Interests.


%


\clearpage

\newpage

\makeatletter 
\renewcommand\onecolumngrid{
\do@columngrid{one}{\@ne}%
\def\set@footnotewidth{\onecolumngrid}
\setlength{\skip\footins}{16pt}
\def\footnoterule{\kern-6pt\hrule width 1.5in\kern6pt}%
}

\renewcommand\twocolumngrid{
        \def\footnoterule{
        \dimen@\skip\footins\divide\dimen@\thr@@
        \kern-\dimen@\hrule width.5in\kern\dimen@}
        \do@columngrid{mlt}{\tw@}
}%
\makeatother    

\renewcommand{\thefigure}{S\arabic{figure}} 
\renewcommand{\theequation}{S\arabic{equation}} 
\renewcommand{\thetable}{S\arabic{table}}
\renewcommand*{\thesubsection}{\Alph{subsection}}

\setcounter{figure}{0} 
\setcounter{equation}{0} 
\setcounter{table}{0}
\setcounter{section}{0}

\title{Supplemental Information for "Exploring possible magnetic monopoles-induced magneto-electricity in spin ices"}

\date{\today}

\maketitle

\onecolumngrid


\section
{Thermally activated processes as a function of off-stoichiometry at high temperature}
\label{SMA}

\begin{figure}
\includegraphics[width=246pt]{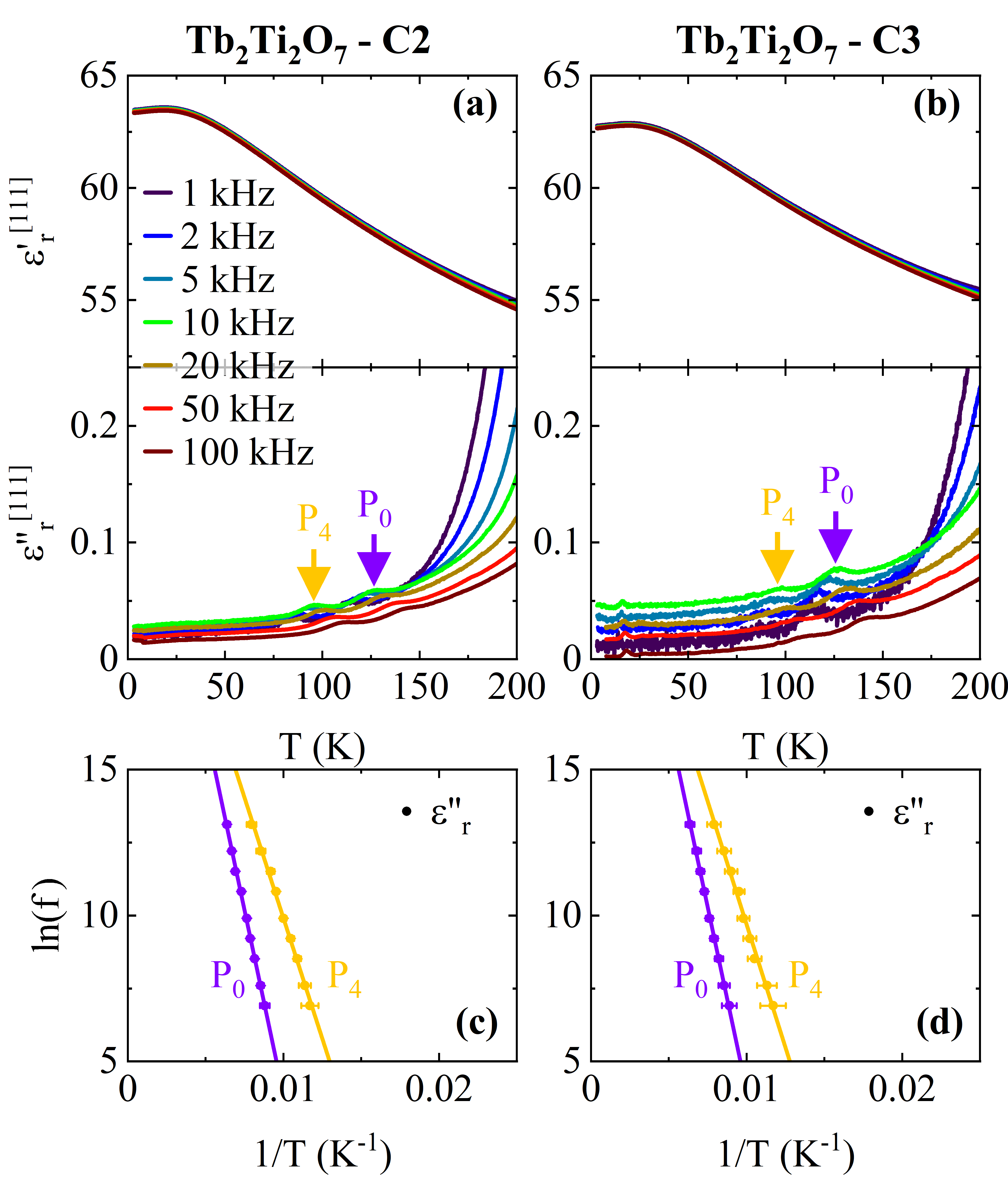}
\caption{Thermally activated processes in two \TTO\ samples, C2 (left) and C3 (right). (a) Real part $\varepsilon'_{\mrm{r}}$ an (b) imaginary part $\varepsilon''_{\mrm{r}}$ of the relative permittivity measured at different frequencies. Two dissipation peaks $P_{0}$ and $P_{4}$ are observed in $\varepsilon''_{\mrm{r}}$. (d) Dissipation peak positions (circles) and fits (lines) from which are extracted the activation energy (slope) and zero-point relaxation time of the relaxation processes.}
\label{fig:SM_Fig1}
\end{figure}

The thermally activated processes observed in the dielectric measurements are dependent on small off-stoichiometry of the samples. This has been checked with two additional \TTO\ samples, C2 and C3 of formula Tb$_{2+x}$Ti$_{2-x}$O$_{7+y}$ with $x=+0.003$ (C2) and $x=-0.003$ (C3) \cite{Alexanian2023}. Two  peaks in the dissipating part of the permittivity are observed, P$_0$ and P$_4$, of lower intensity than in crystal C1 but with similar activated energies ($\SI{1650}{\kelvin}$ and $\SI{2500}{\kelvin}$) and zero point relaxation times ($\SI{3}{\pico\second}$ and $\SI{0.2}{\pico\second}$ respectively). P$_0$ is associated with the same activation process than in crystal C1. Table \ref{tab:SMactivation} gives an overview of the different activation processes observed in the \TTO\ samples as well as the \HTO\ one. 

\begin{table}
\begin{tabular}{l|c|c|c}
\hline
\hline
\noalign{\vskip 0.5mm}
Sample & Process & $E_{a}$ $(\si{\kelvin})$ & $\tau_0$ ($\si{\pico\second}$), $\bm{E}\parallel[111]$ \\
\hline\noalign{\vskip 0.5mm}
\TTO\ - C1 & P$_0$ & 2530 & 0.03 \\
\hline\noalign{\vskip 0.5mm}
\TTO\ - C2 & P$_0$ & 2520 & 0.04 \\
\TTO\ - C2 & P$_4$ & 1650 & 0.53 \\
\hline\noalign{\vskip 0.5mm}
\TTO\ - C3 & P$_0$ & 2510 & 0.03 \\
\TTO\ - C3 & P$_4$ & 1690 & 0.46  \\
\hline\noalign{\vskip 0.5mm}
\HTO & P$_1$ & 5300 & 0.21  \\
\HTO & P$_2$ & 2880 & 0.11 \\
\HTO & P$_3$ & 1750 & 0.28 \\
\hline
\hline
\end{tabular}
\caption{Activation processes observed  in \TTO\ and \HTO\ samples. $E_{a}$ is the activation energy and $\tau_0 = 1/(2\pi f_{0})$ the zero point relaxation time.}
\label{tab:SMactivation}
\end{table}

\section{Electric and magnetic field dependence of the dielectric permittivity in the high temperature regime}
\label{SMB}

\begin{figure}
\includegraphics[width=246pt]{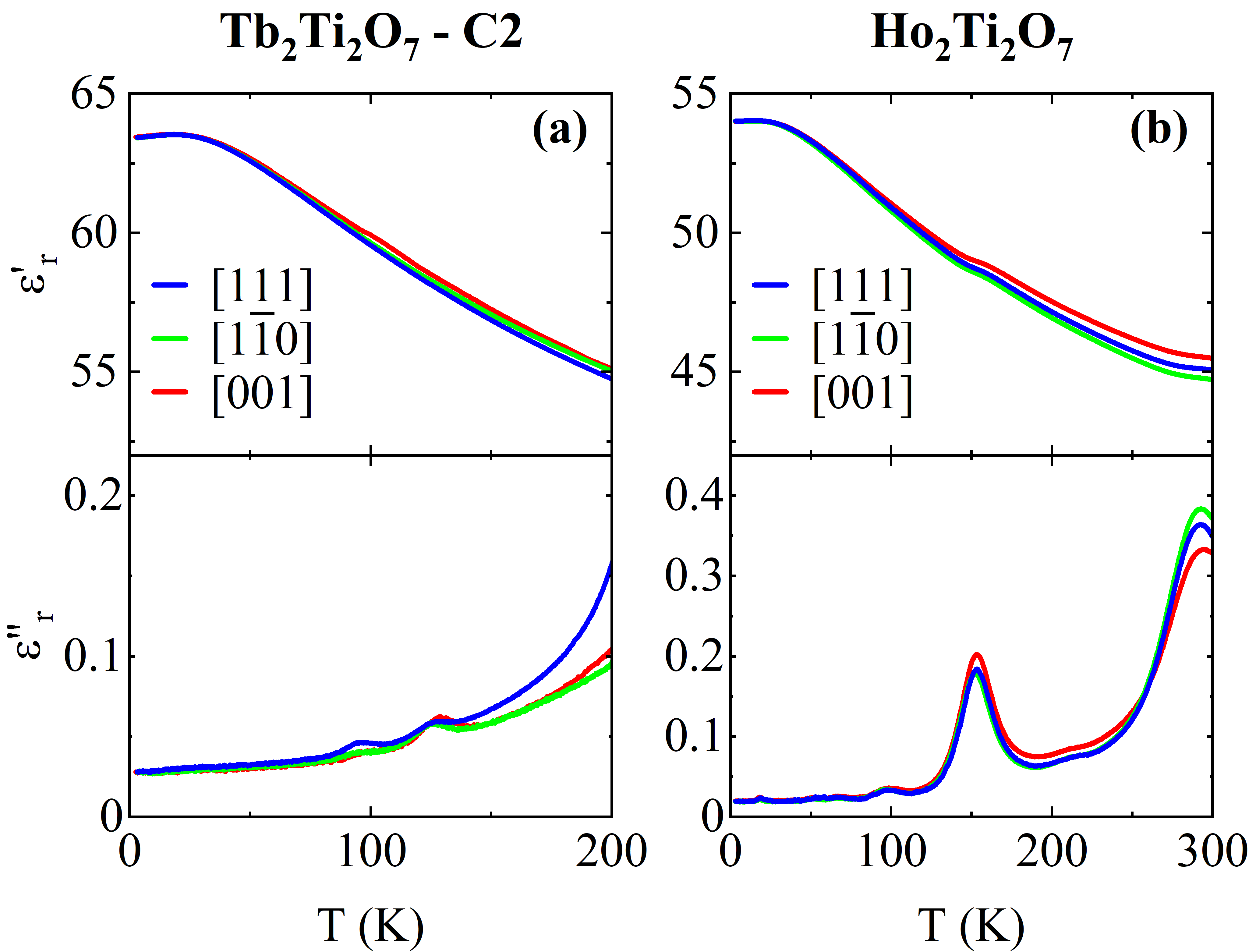 }
\caption{Dependence of the dielectric permittivity on the direction of the electric field in \TTO\ - C2 and \HTO. 3 different plaquettes for each compound were used with $\bm{e}\parallel[111]$, $[1\bar{1}0]$ and $[001]$. The temperature was scanned from $\SI{2.5}{\kelvin}$ up to $\SI{200}{\kelvin}$ for \TTO\ and up to $\SI{300}{\kelvin}$ for \HTO\ with a measurement frequency of $\SI{10}{\kilo\hertz}$.}
\label{fig:SM_Fig2}
\end{figure}

\begin{figure}
\includegraphics[width=246pt]{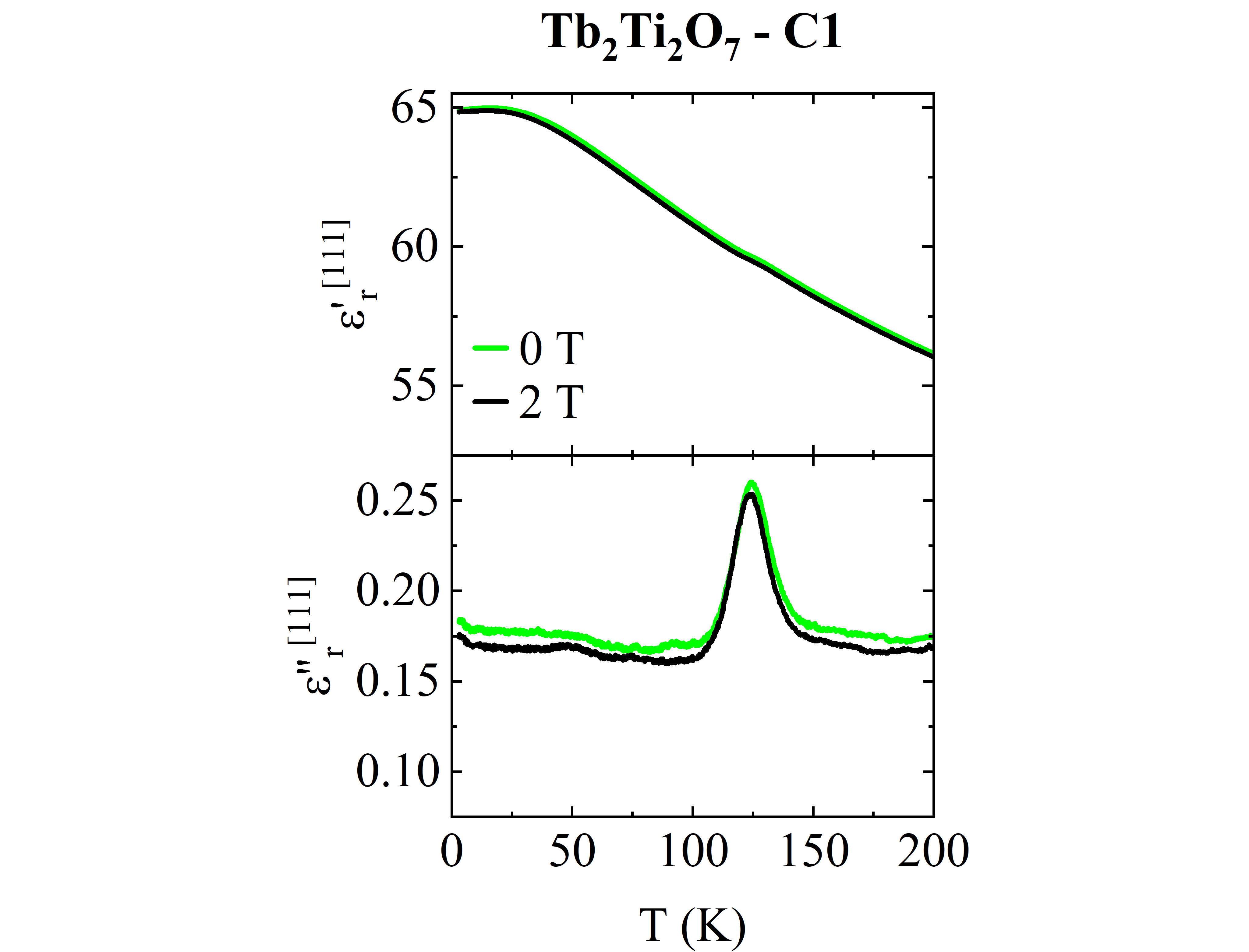 }
\caption{Magnetic field dependence of the dielectric permittivity in \TTO. The real part (top) and imaginary part (bottom) have been measured at $\SI{10}{\kilo\hertz}$ with the AC electric field along $\parallel[111]$ and a static magnetic field of $\SI{0}{\tesla}$ (green) and $\SI{2}{\tesla}$ (black) also along $[111]$.}
\label{fig:SM_Fig3}
\end{figure}

We have explored how the dielectric response depends on the direction of the applied AC electric field. To do this, $\bm{e}$ was applied perpendicular to three different plackets of \HTO\ and \TTO\ along the three principal directions: $[111]$ (3-fold axis), $[001]$ (4-fold axis) and $[1\bar{1}0]$ (2-fold axis). The results presented in Figure \ref{fig:SM_Fig2} do not show any significant dependence on the direction of $\bm{e}$: the slight observed ones are within the uncertainties on the samples size. The activation processes are observed identically: the involved defects respond electrically whatever the direction of the applied electric field. The effect of a $\SI{2}{\tesla}$ static magnetic field has also been investigated in \TTO\ (see Figure \ref{fig:SM_Fig3}): the activation process around $\SI{100}{\kelvin}$ is clearly visible but no change is observed in presence of the magnetic field.

\section{Electric and magnetic field dependence of the pyroelectric current and polarization in the high temperature regime}
\label{SMC}

\begin{figure}
\includegraphics[width=246pt]{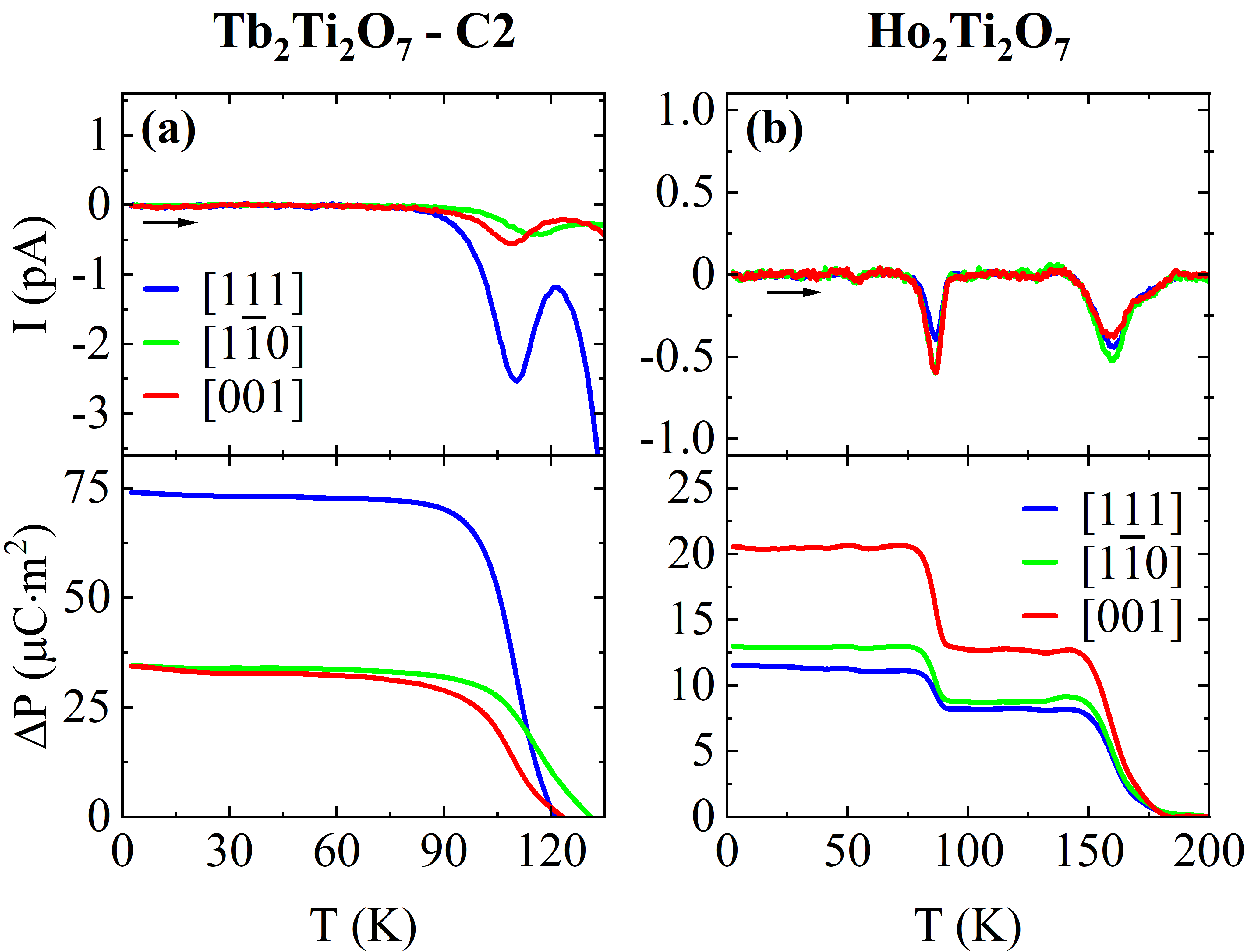 }
\caption{Electric field dependence of the pyroelectric current (top) and electric polarization (bottom) of (a) \TTO\ and (b) \HTO. For \TTO\, the polarization is calculated using a reference at high temperature corresponding to the local minimum in the pyroelectric current around $\SIrange{120}{130}{\kelvin}$.}
\label{fig:SM_Fig4}
\end{figure}

\begin{figure*}[ht!]
\includegraphics[width=1\textwidth]{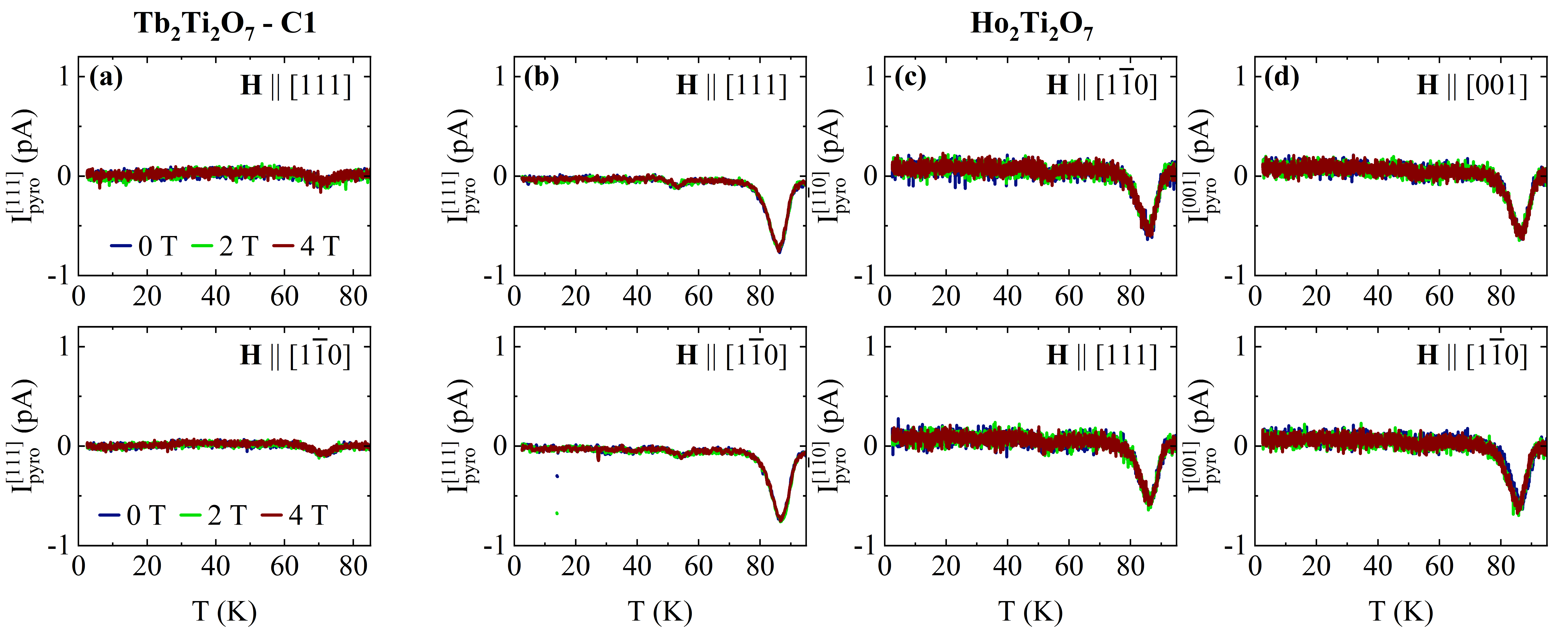 }
\caption{Magnetic field dependence of the pyroelectric current for different samples, in the configuration $\bm{H}\parallel\bm{E}$ (top) or $\bm{H}\perp\bm{E}$ (bottom) : (a) \TTO\ - C1 with $\bm{E}\parallel[111]$ ; (b) \HTO\ with $\bm{E}\parallel[111]$ ; (c) \HTO\ with $\bm{E}\parallel[1\bar{1}0]$ ; (d) \HTO\ with $\bm{E}\parallel[001]$.}
\label{fig:SM_Fig5}
\end{figure*}

As for the dielectric permittivity, we have measured the dependence of the pyroelectric current and of the associated polarization on the electric field direction. We have also probed possible effects of a static magnetic field. The three main directions of measurement for the pyroelectric currents are presented in Figure \ref{fig:SM_Fig4} for \TTO\ crystal $-$ C2 and \HTO. The observed polarization (associated to local defects) is slightly higher along $[111]$ for \TTO\ and $[001]$ for \HTO\ although always of the same order of magnitude.
The effect of a static magnetic field up to 4T has also been investigated either when it is applied in the direction of the current flow or perpendicular to it (see Figure \ref{fig:SM_Fig5}). No effect of the magnetic field is detected confirming that the activated processes observed in \TTO\ and \HTO\ are not affected by a magnetic field.

\begin{figure*}[ht!]
\includegraphics[width=1\textwidth]{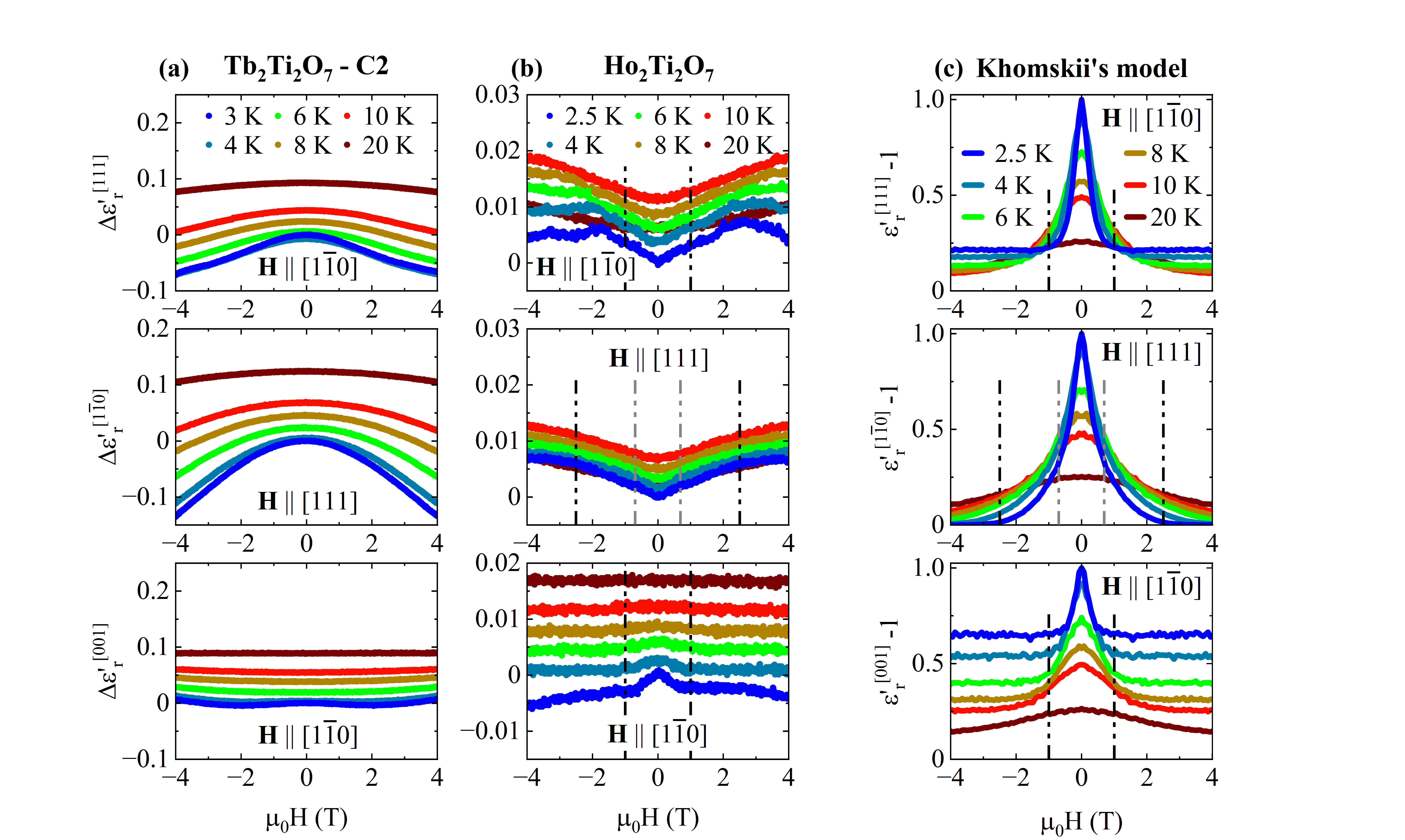 }
\caption{Magnetic field dependence of the dielectric permittivity of (a) \TTO\ and (b) \HTO\ with the electric field $\bm{e}$ applied along the three main directions of the cubic pyrochlore structure and the magnetic field $\bm{H}$ applied perpendicular to it. The reference is taken at zeo field and 3 K, 2.5 K respectively. Note the different scales of $\Delta\varepsilon'_{\mrm{r}}$ for the two compounds. (c) Monte-Carlo calculations of the dielectric susceptibility ($\varepsilon_{\mrm{r}}-1$) using Khomskii's model at the same temperatures and for the same field orientations as the measurements. It is normalized at 1 at 2.5 K and zero field. The vertical dashed lines in (c) point out to the changes of regime of the calculated dielectric susceptibility at 2.5 K.}
\label{fig:SM_Fig6}
\end{figure*}

\section{Magneto-electric response in the low temperature regime in additional fields orientation and additional samples}
\label{SMD}

Measurements of the dielectric permittivity have also been performed in an applied magnetic field $\bm{H}$ perpendicular to the AC electric field $\bm{E}$. The results are presented in Figure \ref{fig:SM_Fig6}. For \TTO\, it is first remarkable that the magnetic field dependence is dominated by a quadratic contribution of the opposite sign to that in the configurations $\bm{E}\parallel\bm{H}$, at least for $\bm{E}\parallel[111]$, $\bm{H}\parallel[1\bar{1}0]$ and vice-versa. For $\bm{E}\parallel[001]$, $\bm{H}\parallel[1\bar{1}0]$, no magnetic field dependence is observed above $\SI{10}{\kelvin}$ and only very small changes are seen at lower temperatures. The situation is again more complex for \HTO\ : while the quadratic contribution mainly visible at $\SI{20}{\kelvin}$ follows the same trend (opposite sign compared to $\bm{E}\parallel\bm{H}$ except for $\bm{e}\parallel[001]$, $\bm{H}\parallel[1\bar{1}0]$ with no quadratic dependence), linear effects are observable below $\SI{8}{\kelvin}$. The Monte Carlo simulations within Komskii's model are mainly dependent on the direction of the magnetic field, with small changes when the AC electric field is rotated from parallel to perpendicular to the magnetic field. The main change occurs for $\bm{H}\parallel[1\bar{1}0]$ where the value of Komskii susceptibility is non zero at $\SI{2.5}{\kelvin}$ and $\SI{4}{\tesla}$. Two out of four magnetic moments are not fixed by the magnetic field in this direction: thermally excited magnetic monopoles and thus Khomskii's electric dipoles can be present even at high magnetic field. These induced electric dipoles are perpendicular to the magnetic field: contrary to the configuration $\bm{E}\parallel\bm{H}$ (see Figure 3 of main text), they can be probe with an electric field $\bm{E}\perp\bm{H}$, resulting in a residual contribution in the dielectric permittivity for $\bm{E}\parallel[111]$ and $\bm{E}\parallel[001]$.

\begin{figure}
\includegraphics[width=246pt]{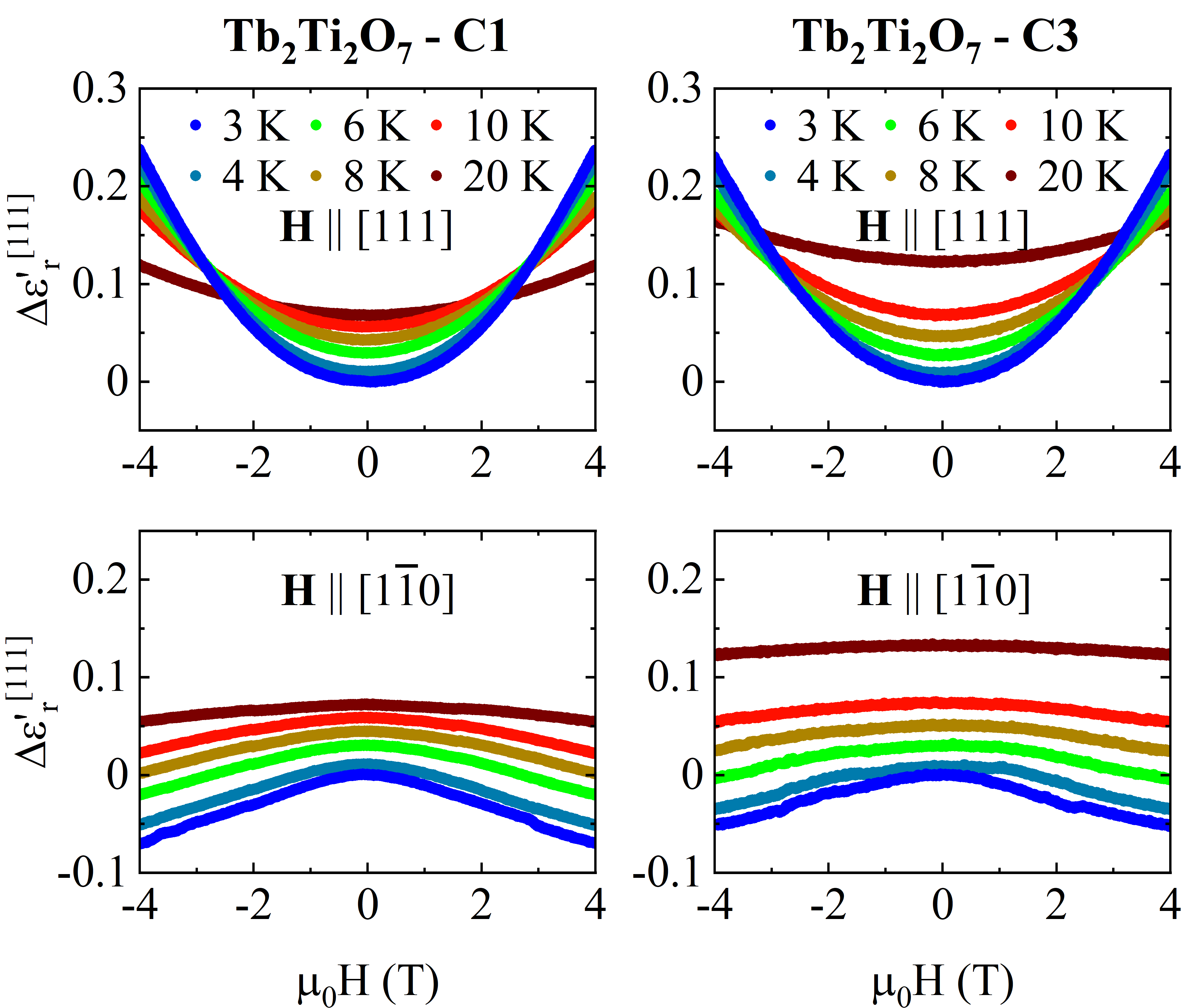 }
\caption{Magnetic field dependence of $\varepsilon'_{r}$ for $\bm{E}\parallel[111]$ of (a) \TTO\ $-$ C1 and (b) \TTO\ $-$ C3. Top pannels : measurements with $\bm{H}\parallel[111]\parallel\bm{E}$. Bottom pannels: measurements with $\bm{H}\parallel\parallel[1\bar{1}0]\perp\bm{E}$.}
\label{fig:SM_Fig7}
\end{figure}

To ensure that the observed magneto-electric effects are reproducible in other samples, magneto-electric measurements were carried out in \TTO\ $-$ C1 and \TTO\ $-$ C3 crystals. A $[111]$ plaquette was used for each sample, alowing measurements in the configurations $\bm{E}\parallel[111]$, $\bm{H}\parallel[111]$ and $\bm{E}\parallel[111]$, $\bm{H}\parallel[1\bar{1}0]$. Results are displayed in Fig. \ref{fig:SM_Fig7}. Even if the temperature evolution of the dielectric permittivity at zero magnetic field is slightly different for the three \TTO\ crystals (which can be due to different thermally activated processes), the magnetic field evolution is similar. 

\section{Magnetic space groups of the different magnetic phases and associated fourth order magneto-dielectric tensors}
\label{SME}

\begin{table*}
\begin{ruledtabular}
\begin{tabular} {l|c|c|c|c}
\noalign{\vskip 0.5mm}
& $\bm{H} = \bm{0}$ & $\bm{H}\parallel[001]$ & $\bm{H}\parallel[1\bar{1}0]$ & $\bm{H}\parallel[111]$  \\
\hline\noalign{\vskip 0.5mm}
Mag. space group & $\mathrm{Fd\bar{3}m1'}$ & $\mathrm{I4_{1}/am'd'}$ & $\mathrm{Imm'a'}$ & $\mathrm{R\bar{3}m'}$ \\
Mag. point group & $\mathrm{m\bar{3}m1'}$ & $\mathrm{4/mm'm'}$ & $\mathrm{mm'm'}$ & $\mathrm{\bar{3}m'}$ \\
\hline\noalign{\vskip 0.5mm}
\TTO, \HTO\ & \hspace{0.24cm} paramagnetic \hspace{0.24cm} & paramagnetic & paramagnetic & paramagnetic \\
\HTO\ & spin ice & \hspace{0.24cm} ordered spin ice \hspace{0.24cm} & \hspace{0.24cm} partially ordered spin ice \hspace{0.24cm} & \hspace{0.24cm} kagome ice, monopole ice \hspace{0.24cm} \\
\hline\noalign{\vskip 0.5mm}
$\Delta^{(2)}\varepsilon'_{r}$ & $H^{2}$ &$H$, $H^{2}$ & $H$, $H^{2}$ & $H$, $H^{2}$ \\
\end{tabular}
\caption{Magnetic space and point groups of the different phases present in \TTO\ and \HTO\ in both ordered ans disordered states. The allowed terms in the magneto-dielectric tensor up to second order are also given.}
\label{tab:groupes_magnétiques}
\end{ruledtabular}
\end{table*}

At high temperature, in the paramagnetic regime, the two pyrochlore compounds are described within the same magnetic space group $\mathrm{Fd\bar{3}m1'}$. Application of a static magnetic field will break some symmetries depending of its orientation with respect to the crystallographic diections (see Table \ref{tab:groupes_magnétiques}). At low temperature, for \HTO\, several magnetic phases are stabilized by the magnetic field \cite{Sazonov2010,Sazonov2012,Sazonov2013}, whose corresponding magnetic space groups are also reported in Table \ref{tab:groupes_magnétiques}. From symmetry arguments, the magneto-dielectric response will always contain a term quadratic with magnetic fields while a linear term is also always allowed except in the magnetic space group $\mathrm{Fd\bar{3}m1'}$.

Anticipating the symmetry analysis of the next section, we reproduce here the symmetry restricted fourth order magneto-dielectric tensors associated to the quadratic response with magnetic field in each magnetic point group \cite{Gallego2019} :

\begin{equation*}
\begin{gathered}
\delta^{\mathrm{m\bar{3}m1'}} = 
\begin{pmatrix}
\delta_{xx} & \delta_{xy} & \delta_{xy} & 0 & 0 & 0 \\
\delta_{xy} & \delta_{xx} & \delta_{xy} & 0 & 0 & 0 \\
\delta_{xy} & \delta_{xy} & \delta_{xx} & 0 & 0 & 0 \\
0 & 0 & 0 & \delta_{44} & 0 & 0 \\
0 & 0 & 0 & 0 & \delta_{44} & 0 \\
0 & 0 & 0 & 0 & 0 & \delta_{44}
\end{pmatrix}\textrm{,}\\
\\
\delta^{\mathrm{4/mm'm'}} = 
\begin{pmatrix}
\Delta_{xx} & \Delta_{xy} & \Delta_{xz} & 0 & 0 & 0 \\
\Delta_{xy} & \Delta_{xx} & \Delta_{xz} & 0 & 0 & 0 \\
\Delta_{zx} & \Delta_{zx} & \Delta_{zz} & 0 & 0 & 0 \\
0 & 0 & 0 & \Delta_{44} & 0 & 0 \\
0 & 0 & 0 & 0 & \Delta_{44} & 0 \\
0 & 0 & 0 & 0 & 0 & \Delta_{66}
\end{pmatrix}\textrm{,}\\
\\
\delta^{\mathrm{mm'm'}} = 
\begin{pmatrix}
d_{xx} & d_{xy} & d_{xz} & 0 & 0 & 0 \\
d_{yx} & d_{yy} & d_{yz} & 0 & 0 & 0 \\
d_{zx} & d_{zy} & d_{zz} & 0 & 0 & 0 \\
0 & 0 & 0 & d_{44} & 0 & 0 \\
0 & 0 & 0 & 0 & d_{55} & 0 \\
0 & 0 & 0 & 0 & 0 & d_{66}
\end{pmatrix}\textrm{,}\\
\\
\delta^{\mathrm{\bar{3}m'}} = 
\begin{pmatrix}
D_{xx} & D_{xy} & D_{xz} & D_{14} & 0 & 0 \\
D_{xy} & D_{xx} & D_{xz} & -D_{14} & 0 & 0 \\
D_{zx} & D_{zx} & D_{zz} & 0 & 0 & 0 \\
D_{41} & -D_{41} & 0 & D_{44} & 0 & 0 \\
0 & 0 & 0 & 0 & D_{44} & 0 \\
0 & 0 & 0 & 0 & 0 & \frac{D_{xx}}{2}-\frac{D_{xy}}{2}
\end{pmatrix}\textrm{.}
\end{gathered}
\label{eq:tenseurs}
\end{equation*}

Note that these symmetry tensors are written with the reduced indices notations. Two symmetry indices $i\in\llbracket 1,3\rrbracket$, $j\in\llbracket 1,3\rrbracket$ are replaced by a unique one $u\in\llbracket 1,6\rrbracket$ so that $u\equiv i$ if $i=j$ and $u\equiv 9 - (i+j)$ if $i\ne j$. They are also written in the natural basis of the considered point group $\left(\bm{a},\bm{b},\bm{c}\right)$. Comparing with the cubic basis $\left(\left[100\right],\left[010\right],\left[001\right]\right)$, we have :
\begin{itemize}
\item For the point group $\mathrm{m\bar{3}m1'}$ ($\bm{H} = \bm{0}$) and $\mathrm{4/mm'm'}$ ($\bm{H}\parallel[001]$) : $\bm{a} = \left[100\right]$, $\bm{b} = \left[010\right]$, $\bm{c} = \left[001\right]$ ;
\item For the point group $\mathrm{mm'm'}$ ($\bm{H}\parallel[1\bar{1}0]$) : $\bm{a} = \left[1\bar{1}0\right]/\sqrt{2}$, $\bm{b} = \left[110\right]/\sqrt{2}$, $\bm{c} = \left[001\right]$ ;
\item For the point group $\mathrm{\bar{3}m'}$ ($\bm{H}\parallel\left[111\right]$) : $\bm{a} = \left[1\bar{1}0\right]/\sqrt{2}$, $\bm{b} = \left[11\bar{2}\right]/\sqrt{6}$, $\bm{c} = \left[111\right]/\sqrt{3}$.
\end{itemize}


\section{Symmetry analysis of the quadratic magnetic field dependence of the dielectric permittivity}
\label{SMF}

The dielectric permittivity tensor $\varepsilon_{ij}$ can be expanded as a function of the magnetic field in the following way:
\begin{equation}
\varepsilon_{ij}(\bm{H}) = \varepsilon_{0}\left(\delta_{ij}+\chi_{ij}^{ee}\right) + \gamma_{ijk}H_{k} + \delta_{ijkl}H_{k}H_{l} + ...
\label{eq:permittivite}
\end{equation}

All these tensors can in principle depend on many external parameters but the temperature dependence is the only one relevant in our experiments. A quadratic term is observed at $\SI{20}{\kelvin}$ and above: in the following we will focus on this term described by the tensor $\delta$, its symmetry restricted expressions being given in \ref{SME} for the different magnetic point groups of spin ices.


To analyze all the measurements in the same framework, one has to assume that we move continuously from the magnetic point group $\mathrm{m\bar{3}m1'}$ to the different magnetic point groups associated to each directions of $\bm{H}$. This is relevant at $\SI{20}{\kelvin}$ and above where the magnetization curves are continuous and change regularly with the applied magnetic field. In this case, one can assume that the components $\Delta_{uv}$, $d_{uv}$ and $D_{uv}$ can be (at least partially) written as a function of the components $\delta_{u'v'}$ of the parent phase associated to the point group $\mathrm{m\bar{3}m1'}$. To do this, we identify the components of these tensors written in the same basis and add a supplementary term $\epsilon$ (assumed weak) accounting for the effect of the electric field $\bm{e}$ and the magnetic field $\bm{H}$ that cannot be captured by the component of the parent phase whose magnetic point group is $\mathrm{m\bar{3}m1'}$. To determine which components are relevant for our measurements where the AC electric field $\bm{e}$ is applied along $[111]$, $[1\bar{1}0]$ or $[001]$ and the static magnetic field $\bm{H}$ is either parallel or perpendicular to it, we use the equation 1 of the main text reproduced here:
\begin{equation}
\Delta\varepsilon'_{\mrm{r}}(H) = (\mu_{0}H)^{2}\sum_{ijkl}\delta_{ijkl}e_{i}e_{j}h_{k}h_{l}.
\end{equation}
In this equation, $e_{i,j}$ ($h_{k,l}$) correspond to the normalized components of the applied electric (magnetic) fields. Results are given in Table \ref{tab:analyse_symetries}.


\begin{table*}
\begin{ruledtabular}
\begin{tabular} {l|c|c}\noalign{\vskip 0.5mm}
$\Delta\varepsilon'_{\mrm{r}}(H)/H^{2}$ & \hspace{3.41cm} $\bm{H}\parallel\bm{E}$ \hspace{3.41cm} & \hspace{3.36cm} $\bm{H}\perp\bm{E}$ \hspace{3.36cm} \\
\hline\noalign{\vskip 0.5mm} 
$\bm{E}\parallel\left[111\right]$& $D_{zz}= \frac{1}{3}\left(4\delta_{44}+\delta_{xx}+2\delta_{xy}\right)+\epsilon_{111}^{\parallel}$ & $\frac{2}{3}d_{yx}+\frac{1}{3}d_{zx}= \frac{1}{3}\left(-2\delta_{44}+\delta_{xx}+2\delta_{xy}\right)+\epsilon_{111}^{\perp}$ \\
$\bm{E}\parallel\left[1\bar{1}0\right]$ & $d_{xx}=\frac{1}{2}\left(2\delta_{44}+\delta_{xx}+\delta_{xy}\right)+\epsilon_{1\bar{1}0}^{\parallel}$ & $D_{xz}= \frac{1}{3}\left(-2\delta_{44}+\delta_{xx}+2\delta_{xy}\right)+\epsilon_{1\bar{1}0}^{\perp}$ \\
$\bm{E}\parallel\left[001\right]$ & $\Delta_{zz}= \delta_{xx} + \epsilon_{001}^{\parallel}$ & $d_{zx} = \delta_{xy}+\epsilon_{001}^{\perp}$ \\
\end{tabular}
\end{ruledtabular}
\caption{Magneto-electric tensor components measured experimentally and their expression in the cubic $\mathrm{m\bar{3}m1'}$ magnetic point group. The static magnetic field  $H$ was applied either parallel or perpendicular to the AC electric field $E$. For $\bm{H}\perp\bm{E}$, the direction of the magnetic field is $\bm{H}\parallel\left[1\bar{1}0\right]$ (for $\bm{E}\parallel\left[111\right]$ and $\bm{E}\parallel\left[001\right]$) and $\bm{H}\parallel\left[111\right]$ (for $\bm{E}\parallel\left[1\bar{1}0\right]$).}
\label{tab:analyse_symetries}
\end{table*}

\begin{table*}
\begin{ruledtabular}
\begin{tabular}{c|c|c|c|c|c|}
\multicolumn{2}{c|}{} & \multicolumn{2}{c|}{\TTO\ } & \multicolumn{2}{c|}{\HTO\ } \\
\hline\noalign{\vskip 0.5mm} 
\hspace{0.2cm} $\bm{e}$ \hspace{0.2cm} & \hspace{0.2cm} $\bm{H}$ \hspace{0.2cm} & \hspace{0.825cm} $a$ ($\SI{e-3}{\tesla^{-2}}$) \hspace{0.825cm} & \hspace{1.561cm} R$^{2}$ \hspace{1.561cm} & \hspace{0.825cm} $a$ ($\SI{e-3}{\tesla^{-2}}$) \hspace{0.825cm} & \hspace{1.561cm} R$^{2}$ \hspace{1.561cm} \\
\hline\noalign{\vskip 0.5mm} 
$[111]$ & $[111]$ & $3.056\pm0.003$ & 0.999 & $-0.147\pm0.003$  & 0.993\\
$[1\bar{1}0]$ & $[1\bar{1}0]$ & $2.600\pm0.005$ & 0.999 & $-0.221\pm0.006$  & 0.868\\
$[001]$ & $[001]$ & $1.117\pm0.003$ & 0.999 & $0.534\pm0.005$  & 0.999\\
$[111]$ & $[1\bar{1}0]$ & $-1.025\pm0.003$ & 0.999 & $0.294\pm0.005$  & 0.998\\
$[1\bar{1}0]$ & $[111]$ & $-1.192\pm0.003$ & 0.999 & $0.328\pm0.004$  & 0.997\\
$[001]$ & $[1\bar{1}0]$ & & & &\\
\end{tabular}
\end{ruledtabular}
\caption{Fit of the magnetic field dependence of the dielectric permittivity at $T=\SI{20}{\kelvin}$ with a function $a(\mu_{0}H)^{2}+b$, up to $\SI{4}{\tesla}$ for \TTO\ and $\SI{3}{\tesla}$ for \HTO.}
\label{tab:1}
\end{table*}


To go further, we fit all the $T=\SI{20}{\kelvin}$ data measured in $\bm{E}\parallel\bm{H}$ (see Fig. 3 of main text) and $\bm{e}\perp\bm{H}$ (see Fig. \ref{fig:SM_Fig7}) configurations for both compounds to get the coefficient of the quadratic dependence of $\Delta\varepsilon'_{\mrm{r}}(H)$ with magnetic field. Results are shown in Table \ref{tab:1} This allows additional simplifications, especially when considering the measurements in configurations with $\bm{H}\perp\bm{E}$ :
\begin{enumerate}
\item  $\bm{E}\parallel\left[001\right]$ and $\bm{H}\parallel\left[1\bar{1}0\right]$.  No dependence on magnetic field is observed for for \TTO\ and \HTO. Thus, it is reasonable to assume that $\delta_{xy} = \epsilon_{001}^{\perp} = 0$.
\item $\bm{E}\parallel\left[111\right]$, $\bm{H}\parallel\left[1\bar{1}0\right]$ and $\bm{E}\parallel\left[1\bar{1}0\right]$, $\bm{H}\parallel\left[111\right]$. The magnetic field dependence is nearly identical for these two fields orientations, although different in \TTO\ and \HTO. Thus $ \left(-2\delta_{44}+\delta_{xx}\right)/3 + \epsilon_{111}^{\perp} = \left(-2\delta_{44}+\delta_{xx}\right)/3 + \epsilon_{1\bar{1}0}^{\perp}$ so one has $\epsilon_{111}^{\perp} = \epsilon_{1\bar{1}0}^{\perp}$ and it is tempting to assume $\epsilon_{111}^{\perp} = \epsilon_{1\bar{1}0}^{\perp} = 0$.
\end{enumerate}

Finally, only the quantities $\delta_{xx}$, $\delta_{44}$ and $\epsilon_{e}^{\parallel}$ seems to be relevant in the high temperature measurements. Excluding for a moment all the quantities $\epsilon^{\parallel}$, it is possible to get a simplified expression of the dielectric permittivity as a function of magnetic field when its dependence is quadratic : 
\begin{equation}
\begin{gathered}
\varepsilon'_{r}\left(\bm{H}\parallel\left[001\right]\right) = 
H^{2}\begin{pmatrix}
0 & 0 & 0 \\
0 & 0 & 0 \\
0 & 0 & \delta_{xx}
\end{pmatrix}\textrm{,}\\
\varepsilon'_{r}\left(\bm{H}\parallel\left[1\bar{1}0\right]\right) = 
\frac{H^{2}}{2}\begin{pmatrix}
\delta_{xx} & -2\delta_{44} & 0 \\
-2\delta_{44} & \delta_{xx} & 0 \\
0 & 0 & 0
\end{pmatrix}\textrm{,}\\
\varepsilon'_{r}\left(\bm{H}\parallel\left[111\right]\right) = 
\frac{H^{2}}{3}\begin{pmatrix}
\delta_{xx} & 2\delta_{44} & 2\delta_{44} \\
2\delta_{44} & \delta_{xx} & 2\delta_{44} \\
2\delta_{44} & 2\delta_{44} & \delta_{xx} \\
\end{pmatrix}\textrm{.}
\end{gathered}
\label{eq:tenseurs_eps}
\end{equation}

It is also possible to determine the sign of the coefficients $\delta_{xx}$ and $\delta_{44}$ for the two compounds. An inspection of Table \ref{tab:analyse_symetries} combined to Table \ref{tab:1} for the configuration $\bm{e}\parallel\bm{H}\parallel[001]$ show that $\delta_{xx}$ has to be positive for both compounds. One also finds that the coefficients $\delta_{44}$ should be of opposite sign for the two compounds, $\delta_{44}^{\,\mrm{Tb}} > 0$, $\delta_{44}^{\,\mrm{Ho}} < 0$, and should respect the condition $\left|2\delta_{44}\right|>\delta_{xx}$. Moreover, it is possible to obtain a remarakable agreement between the expressions of $\Delta^{(2)}\varepsilon'_{r}$ of Table \ref{tab:analyse_symetries} and the experimental coefficients of the quadratic dependence of $\Delta\varepsilon'_{\mrm{r}}(H)$ with magnetic field ($a$ in Table \ref{tab:1}) for \TTO. With $\delta_{xx} = \SI{1.1e-3}{\tesla^{-2}}$ and $\delta_{44} = \SI{2.0e-3}{\tesla^{-2}}$, one gets:
\begin{itemize}
\item $\Delta^{(2)}\varepsilon'_{r} = \SI{3.0e-3}{\tesla^{-2}}$ for $\bm{e}\parallel\bm{H}\parallel\left[111\right]$ ;
\item $\Delta^{(2)}\varepsilon'_{r} = \SI{2.6e-3}{\tesla^{-2}}$ for $\bm{e}\parallel\bm{H}\parallel\left[1\bar{1}0\right]$ ;
\item $\Delta^{(2)}\varepsilon'_{r} = \SI{1.1e-3}{\tesla^{-2}}$ for $\bm{e}\parallel\bm{H}\parallel\left[001\right]$ ;
\item $\Delta^{(2)}\varepsilon'_{r} = \SI{-1.0e-3}{\tesla^{-2}}$ for $\bm{e}\parallel\left[111\right]$, $\bm{H}\parallel\left[1\bar{1}0\right]$ and $\bm{e}\parallel\left[1\bar{1}0\right]$, $\bm{H}\parallel\left[111\right]$.
\end{itemize}
Such an agreement is impossible to get for \HTO, showing that the quantities $\epsilon^{\parallel}$ cannot be neglected. Indeed, setting $\delta_{xx}= \SI{0.53e-3}{\tesla^{-2}}$ (the fit value of the measurement with $\bm{E}\parallel\bm{H}\parallel[001]$, see Table \ref{tab:1}) lead to quite different $\delta_{44}$ coefficients depending on which $(\bm{E}$,$\bm{H})$ measurement configuration is choosen to calculate it: $-\SI{0.17e-3}{\tesla^{-2}}$ ($\bm{E}\parallel[111], \bm{H}\parallel[1\bar{1}0]$), $-\SI{0.23e-3}{\tesla^{-2}}$ ($\bm{E}\parallel[1\bar{1}0],\bm{H}\parallel[111]$), $-\SI{0.24e-3}{\tesla^{-2}}$ ($\bm{E}\parallel\bm{H}\parallel[111]$) or $-\SI{0.49e-3}{\tesla^{-2}}$ ($\bm{E}\parallel\bm{H}\parallel[1\bar{1}0]$). This could be due to the fact that spins are strongly Ising for \HTO\ but only soft Ising for \TTO, allowing to have smooth variations with the magnetic field.

\end{document}